\documentclass{IEEEtran}
\normalsize

\ifCLASSINFOpdf
\else
\fi

\usepackage{comment} 

\usepackage{cite}
\usepackage{amsmath,amssymb,amsfonts}
\usepackage{textcomp}
\usepackage{bm}
\usepackage[pdftex]{graphicx}
\usepackage[noend]{algorithmic}
\usepackage{algorithm}
\usepackage{multicol}
\usepackage{lipsum, color}
\usepackage{mathtools, cuted, nccmath}
\usepackage{commath}

\usepackage{epstopdf}
\usepackage{subfigure}
\usepackage[short]{optidef}
\usepackage[utf8]{inputenc}
\usepackage{epsfig}
\DeclareGraphicsExtensions{.eps}
\usepackage{multirow} 

\usepackage{subfig}
\usepackage{physics}

\DeclareMathOperator*{\argmin}{argmin}
\DeclareMathOperator*{\argmax}{argmax}

\usepackage{xcolor, soul}
\sethlcolor{red}

\usepackage [english]{babel}
\usepackage [autostyle, english = american]{csquotes}
\MakeOuterQuote{"}

\begin{document}
\title{{\LARGE Reconfigurable Intelligent Surface (RIS)-Enhanced Two-Way OFDM Communications}}
\author{Chandan~Pradhan,~\IEEEmembership{Student Member,~IEEE,}
       Ang~Li,~\IEEEmembership{Member,~IEEE,}
       Lingyang~Song,~\IEEEmembership{Fellow,~IEEE,}
       Jun Li,~\IEEEmembership{Senior Member,~IEEE,}
       Branka~Vucetic,~\IEEEmembership{Fellow,~IEEE,}
       and~Yonghui~Li,~\IEEEmembership{Fellow,~IEEE \vspace{-2.95em}}
    


\thanks{Chandan Pradhan, Yonghui Li and Branka Vucetic are  with  the  Centre  of  Excellence  in  Telecommunications, School of Electrical and Information Engineering, University of Sydney, Sydney, NSW 2006, Australia. (e-mail: \{chandan.pradhan, yonghui.li, branka.vucetic\}@sydney.edu.au).}
\thanks{Ang Li is with the Faculty of Electronic and Information Engineering, Xi’an Jiaotong University, Xi’an 710049, China. (e-mail: ang.li.2020@xjtu.edu.cn).}
\thanks{Lingyang Song is with Peking University, Beijing 100871, China (email: lingyang.song@pku.edu.cn).}
\thanks{Jun Li is with Nanjing University of Science and Technology, Nanjing 210094, China, (email: jun.li@njust.edu.cn).}
}


\IEEEaftertitletext{\vspace{-0.0\baselineskip}}

\maketitle

\begin{abstract}

In this paper, we focus on the reconfigurable intelligent surface (RIS)-enhanced two-way device-to-device (D2D) multi-pair orthogonal-frequency-division-multiplexing (OFDM) communication systems. Specifically, we maximize the minimum bidirectional weighted sum-rate by jointly optimizing the sub-band allocation, the power allocation and the discrete phase shift (PS) design at the RIS. To tackle the main difficulty of the non-convex PS design at the RIS, we firstly formulate a semi-definite relaxation problem and further devise a low-complexity solution for the PS design by leveraging the projected sub-gradient method. We demonstrate the desirable performance gain for the proposed designs through numerical results.

\end{abstract}

\vspace{-1.0mm}

\begin{IEEEkeywords}

Two-way communications, reconfigurable intelligent surfaces (RISs), OFDM.

\end{IEEEkeywords}

\vspace{-4.5mm}

\section{Introduction}
\vspace{-1mm}

With the recent advances in electromagnetic (EM) meta-surfaces, the reconfigurable intelligent surfaces (RISs) are foreseen as the cost-effective and energy-efficient substitutes for the relay-based systems \cite{Zhang2019APB, di2019hybrid}. The RIS is a planar array consisting of a large number of reflecting elements, which are implemented with low-cost programmable positive-intrinsic-negative (PIN) diodes or phase shifters (PSTs) \cite{ruiZhangMag2019, di2019hybrid}. Accordingly, the reflection behaviour of the impinging EM signals can be controlled through an appropriate design of phase shifts (PSs) of the reflecting elements to improve the performance of a wireless network. Compared to the relay-based systems, the deployment of RIS does not involve additional RF chains and the imposition of thermal noise. Furthermore, the RIS can readily be fabricated in small size and low weight, which can be coated on the buildings' facade, walls, etc \cite{ruiZhangMag2019}.


Due to the above benefits RISs provide, they are envisioned to enhance the performance for various wireless applications, by improving the spectral and energy efficiencies, facilitating the simultaneous wireless and power transfer, the massive device-to-device (D2D) communications, etc., \cite{ruiZhangMag2019, di2019hybrid}. While the majority of current works for the RIS-enhanced communication systems focus on one-way communication, there are only a limited number of works that consider the RIS-enhanced two-way communication systems \cite{zhangTwoWay2020, atapattu2020reconfigurable}. The  works in \cite{zhangTwoWay2020, atapattu2020reconfigurable} are primarily limited to a single pair of full-duplex (FD) nodes, where the benefit of two–way network is conditioned on the proper self-interference (SI) cancellation at the FD nodes. This comes at the cost of high hardware complexity and low energy efficiency \cite{bharadia2013full}, which is hence unsuitable for low-cost and power-limited nodes. Furthermore, to the best of our knowledge, the use of orthogonal-frequency-division-multiplexing (OFDM) for the RIS-enhanced two-way communications has not been explored yet.




Motivated by this, we focus on a RIS-enhanced two-way D2D communication system where multiple pairs of transceiver nodes communicate bidirectionally via RIS through the OFDM. Specifically, the available bandwidth is divided into multiple orthogonal sub-bands, where each of the bidirectional communication links across multiple node pairs is allocated  a  subset of non-overlapping sub-bands. We aim to maximize the minimum bidirectional weighted sum-rate by jointly optimizing the sub-band allocation, the power allocation and the PSs at the RIS. We consider the practical discrete PSs at each of its reflecting elements. To tackle the main difficulty of the non-convex  PS design at the RIS, we firstly propose a semi-definite relaxation (SDR) formulation. Subsequently, we devise a low-complexity solution for the PS design by leveraging the projected sub-gradient (PSG) method to achieve a more favorable performance-complexity tradeoff. Numerical results reveal the desirable performance gain for the proposed designs. 


\textit{Notations}: $y$, ${\bf y}$ and $\mathbf{Y}$ denote scalar, vector and matrix, respectively; Conjugate, transpose and conjugate transpose operators are represented by $\left(\cdot\right)^*$, $\left(\cdot\right)^T$ and $\left(\cdot\right)^H$, respectively; $\norm{\cdot}_2$ denotes the $\ell_2$ norm; ${\rm Tr}\left\{\cdot\right\}$ denotes the trace operator; $\ast$ denotes the convolution operation; Expectation of a random variable is noted by $\mathbb{E}[\cdot]$; ${\Re}$ denotes the real part of a complex number; $|\cdot|$ and $\angle$ return the absolute value and the argument of a complex number, respectively. 


\begin{figure}
\centering
\includegraphics[scale=0.24]{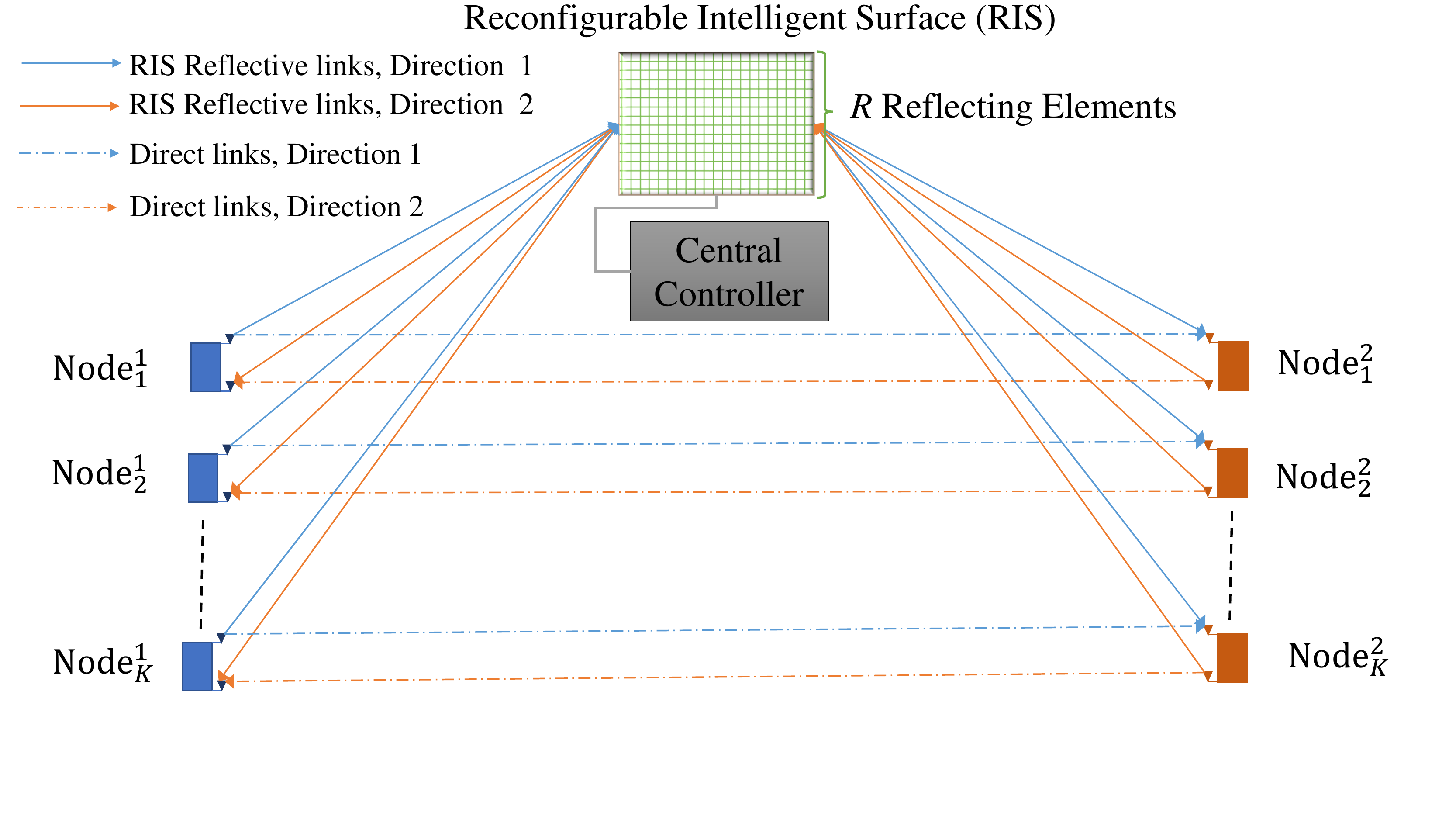}
\vspace{-6mm}
  \caption{\small{RIS-enhanced two-way D2D communication system.}}
\vspace*{-6mm}
\end{figure}
\vspace{-4mm}

\section{System Model}

We consider a two-way D2D communication system with $K$ node pairs, denoted by ${\rm Node}_k^1- {\rm Node}_k^2, k = \left\{1, \cdots K \right\}$, where each node is equipped with single transmit and receive antennas as shown in Fig.1. A RIS with $R$ reflecting elements is deployed to enhance the communication for the $K$ node pairs. Accordingly, let ${\bm \Psi} = \left[\Psi_1, \cdots, \Psi_R \right]^T \in \mathbb{C}^{R \times 1}$ denote the vector of the reflection coefficients at the RIS, such that each reflection coefficient $\Psi_r$ satisfies $\Psi_r \in \mathcal{R} \triangleq \left\{ e^{\jmath \frac{2 \pi b}{2^B}} \big| b = 0, \cdots, 2^B-1  \right\},  \forall r$, where $B$ is the number of quantization bits of the PSTs \cite{zhangTwoWay2020}. Following OFDM, the available bandwidth is divided into $V$ sub-bands, and the sub-bands are occupied to carry out simultaneous communication for the $K$ node pairs, where each node pair further adopts bidirectional communication\footnote{\hspace{-0.1mm}Note that in our work, we denote the transmission from ${\rm Node}_k^1$ to ${\rm Node}_k^2, \forall k$ and vice-versa as direction $1$ and $2$, respectively.} through non-overlapping sub-bands to avoid SI. Accordingly, the communication within each node pair takes through two links, i.e., the ${\rm Node}_k^1$-RIS-${\rm Node}_k^2$ reflected link and the direct ${\rm Node}_k^1$-${\rm Node}_k^2$ link, along both the directions.


\vspace{-4mm}
\subsection{Channel Model}

For the $k$-th node pair, the time-domain ${\rm Node}_k^1$-RIS-${\rm Node}_k^2$ reflected channel via each $r$-th reflection element of the RIS, in the $i$-th direction, is the convolution of the  ${\rm Node}_k^1$-RIS channel, the RIS reflection coefficient, and the RIS-${\rm Node}_k^2$ channel, i.e.,  ${\bm \xi}_{k-r}^i \ast \Psi_r \ast {\bm \xi}_{r-k}^i = \Psi_r {\bm \xi}_{k-r}^i \ast {\bm \xi}_{r-k}^i, i \in \left\{1,2\right\}$, where ${\bm \xi}_{k-r}^i \in \mathbb{C}^{ L_{k-r} \times 1}$ and ${\bm \xi}_{r-k}^i \in \mathbb{C}^{L_{r-k} \times 1}$ denote the time-domain ${\rm Node}_k^1$-RIS and RIS-${\rm Node}_k^2$ channels, respectively, and $L_{k-r}$ and $ L_{r-k}$ are the number of the corresponding delay taps \cite{ruiZhangOfdma2020, zhengOFDMIRS2020}. Similarly, let ${\bm \xi}^i_{k-k} \in \mathbb{C}^{L_{k-k} \times 1}$ denote the time-domain direct ${\rm Node}_k^1$-${\rm Node}_k^2$ channel, where $L_{k-k}$ is the number of delay taps. Furthermore, for each multi-path channel, the channel taps are assumed to follow the exponential power-delay feature characterized by $\left[{\bm \xi}_{x}^i\right]_l = \sqrt{\varrho_x \frac{1 - \alpha}{1 - \alpha^{L_{x}}}} \alpha^{l/2} {\nu}^i_{x_l},  \forall l = 0, \cdots, L_{x} - 1, x \in \left\{k-k, k-r, r-k\right\}$, where $0 < \alpha < 1$, $\varrho_x$ is the path loss and $\nu^i_{x_l} \sim \mathcal{CN}(0, 1)$ represents the small scale fading \cite{analysis2008}.

Accordingly, for the $k$-th node pair, the zero-padded concatenated ${\rm Node}_k^1$-RIS  and RIS-${\rm Node}_k^2$ time-domain channel through each of the $r$-th reflecting element in the $i$-th direction is given  by $\tilde{\bf h}_{k,r}^i = \left[\left({\bm \xi}_{k-r}^i \ast {\bm \xi}_{r-k}^i \right)^T, 0, \cdots, 0 \right]^T \in \mathbb{C}^{V \times 1}$ and $\tilde{\bf H}_k^i = \left[\tilde{\bf h}_{k,1}^i, \cdots, \tilde{\bf h}_{k,R}^i \right]$ \cite{ruiZhangOfdma2020}. Thus, the composite ${\rm Node}_k^1$-RIS-${\rm Node}_k^2$ reflected channel can be expressed as $\tilde{\bf H}_k^i {\bm \Psi}, i \in \left\{1,2\right\}$. By further denoting $\tilde{\bf g}_{k}^i = \left[{\bm \xi}_{k-k}^i, 0, \cdots,0 \right]^T \in \mathbb{C}^{V \times 1}$ as the zero-padded time-domain ${\rm Node}_k^1$-${\rm Node}_k^2$ direct channel, the effective channel impulse response of the $k$-th node pair in the $i$-th direction is given by $ \tilde{\bf h}_k^i = \tilde{\bf g}_{k}^i + \tilde{\bf H}_k^i {\bm \Psi}$. We further assume that the inter-symbol interference is perfectly eliminated through the use of the cyclic prefix as in \cite{ruiZhangOfdma2020}. Finally, for the $k$-th node pair, the channel frequency response on the $v$-th sub-band in the $i$-th direction is expressed as $ \bar{h}_{k_v}^i =  {\bf f}^H_v \tilde{\bf h}_k^i = {\bf f}^H_v \tilde{\bf g}_k^i  + {\bf f}^H_v \tilde{\bf H}_k^i {\bm \Psi} =  g_{k_v}^i + \left({\bf h}_{k_v}^i\right)^H {\bm \Psi}$, where ${\bf f}^H_v$ denotes the $v$-th row of the discrete Fourier transform (DFT) matrix ${\bf F} \in \mathbb{C}^{V \times V}$, $g_{k_v}^i \triangleq {\bf f}^H_v \tilde{\bf g}_k^i$ and ${\bf h}^i_{k_v} \triangleq \left(\tilde{\bf H}_k^i\right)^H {\bf f}_v$. Furthermore, the RIS is considered to be attached with a central controller, which controls the PSs of its reflecting elements and communicates with the node pairs via dedicated wireless links for coordinating transmission and exchanging information on the channel state information (CSI) \cite{zhengOFDMIRS2020, zhang2020joint}, where the perfect CSI is assumed to be estimated and known at the nodes through the central controller as in \cite{zhengOFDMIRS2020, Zhang2019APB,  zhangTwoWay2020}. Based  on  the  CSI, the central controller performs the sub-band and power allocations as well as the PS design at the RIS.


\vspace{-5mm}
\subsection{Transmission Model}

 To avoid the inter-node and inter-directional interferences, each sub-band is allocated to at most one ${\rm Node}_k^i$. Accordingly, let $\eta_{k_v}^i$ indicate whether the $v$-th sub-band is allocated to ${\rm Node}_k^i, i \in \left\{ 1,2\right\}$, i.e., $\eta^i_{k_v} = 1$ if $v$-th sub-band is assigned to  ${\rm Node}_k^i$, and $\eta^i_{k_v} = 0$ otherwise. Thus, we have $\sum_{k=1}^K \sum_{i=1}^2 \eta_{k_v}^i = 1, \forall v$. Moreover, we consider a total transmit power constraint for each ${\rm Node}_k^i$, which is given by $\sum_{v = 1}^{V} \eta_{k_v}^i {p}^i_{k_v} \leq P_{k}^i, \forall k, i \in \left\{1,2 \right\}$, where $p_{k_v}^i \geq 0$ denotes the transmit power allocated to $v$-th sub-band by ${\rm Node}_k^i$ and $P_k^i$ is the maximum power at ${\rm Node}_k^i$. Accordingly, the received signal on the $v$-th sub-band at ${\rm Node}_k^{\bar{i}}, \bar{i} \in \left\{1,2\right\}\backslash\left\{i\right\}$ in the $i$-th direction, when $\eta_{k_v}^i = 1$, is given by $y_{k_v}^i =  \sqrt{p_{k_v}^i}\bar{h}_{k_v}^i s_{k_v}^i + z_{k_v}^i$, where  $s_{k_v}^i$ is the transmitted signal such that  $\mathbb{E}\left[|s_{k_v}^i|^2\right] = 1$ and $z_{k_v}^i \sim \mathcal{CN}(0, \sigma_{k_v}^2)$ denotes the additive white Gaussian noise (AWGN) in the sub-band. Subsequently, the achievable rate in bits per second per Hertz (bps/Hz) for the $k$-th node pair on the $v$-th sub-band in the $i$-th direction is given by $\Gamma_{k_v}^i = \frac{\eta_{k_v}^i}{V} \log_2 \left(1 + \gamma_{k_v}^i  \right)$, where $\gamma_{k_v}^i \triangleq \frac{p_{k_v}^i \left|\bar{h}_{k_v}^i\right|^2}{ \sigma_{k_v}^2}$ \cite{ruiZhangOfdma2020}.

\vspace{-4mm}

\section{Problem Formulation and Proposed Solution}

In this work, we aim to maximize the minimum bidirectional weighted sum-rate across all the sub-bands for the $K$ node pairs. Accordingly, we formulate the following max-min optimization problem:

\vspace{-3mm}

\begin{equation}
\begin{aligned}
& \mathcal{P}_1: && \underset{\left\{{\bm \eta}_1, {\bm \eta}_2, {\bf p}_1, {\bf p}_2, {\bm \Psi}\right\}}{\max}\underset{\left\{i \in \{1,2\}\right\}} \min \; \sum_{k=1}^K  \sum_{v = 1}^{V} \varkappa_k \Gamma_{k_v}^i\\
& \text{\it s.t.}
& &  {\rm C}_1: \sum_{v = 1}^{V} \eta_{k_v}^i {p}^i_{k_v} \leq P_{k}^i, \forall k, i \in \left\{1,2 \right\}, \\
&&& {\rm C}_2: {p}^i_{k_v} \geq 0, \forall k,v, i \in \left\{1,2 \right\},\\
&&& {\rm C}_5:  \sum_{k=1}^K \sum_{i=1}^2 \eta_{k_v}^i = 1, \forall v, \\
&&& {\rm C}_6: \eta_{k_v}^i = \left\{0,1\right\}, \forall k, v, i \in \left\{1,2\right\},\\
&&& {\rm C}_7: {\Psi}_r \in \mathcal{R}, \; \forall r. 
\end{aligned}
\end{equation}
where $\varkappa_k$ is the weighting factor of the $k$-th node pair, ${\bm \eta}_i \triangleq \left[\eta_{1_1}^i, \cdots,  \eta_{K_V}^i \right]^T$ and ${\bf p}_i \triangleq \left[p_{1_1}^i, \cdots,  p_{K_V}^i \right]^T, i \in \left\{ 1,2\right\}$. Note that along with the non-convex constraints ${\rm C}_6$ and ${\rm C}_7$, the coupling of $\Gamma_{k_v}^i, \forall k,v, i \in \left\{1,2\right\}$ through ${\bm \Psi}$ makes $\mathcal{P}_1$ difficult to solve. Accordingly, to  solve $\mathcal{P}_1$,  we  present  a  two-stage design, which is described in the subsequent sub-sections.


\vspace{-5mm}

\subsection{First-Stage: Sub-Band Allocation}

Firstly, it can be observed from $\mathcal{P}_1$ that the sub-band allocation problem, for a given $\left\{{\bm \Psi}, {\bf p}_i, i \in \left\{1,2\right\}\right\}$, is a non-convex binary integer problem due to ${\rm C}_6$. Accordingly, to reduce the computational complexity for the sub-band allocation problem, we resort to a sub-optimal algorithm, given in Algorithm 1, similar to that discussed in \cite{rhee2000}. Specifically, we first obtain an appropriate initial ${\bm \Psi}$, denoted by $\bar{\bm \Psi}$, which maximizes the minimum bidirectional effective channel gain across all the sub-bands for the $K$ node pairs, as detailed in Appendix A. Subsequently, assuming uniform power allocation across the sub-bands, the node with the lowest sum-rate across both the directions is iteratively assigned a sub-band where it achieves the highest rate $\Gamma_{k_v}^i$, as described in Algorithm 1, where $\Gamma_{k_v}^i \triangleq \log_2\left(1 + \gamma_{k_v}^i|_{{\bm \Psi}={\bm \bar{\bm \Psi}}} \right), \forall k, v, i \in \left\{1,2 \right\}$ and $\mathcal{V}$ is defined as the set of available sub-bands.

%

\vspace{-2mm}

\setlength{\textfloatsep}{10pt}

\begin{algorithm}[!htb]
 \begin{algorithmic}[1]
\STATE \textbf{Input}: Initial ${\bm \Psi} =\bar{\bm \Psi}$ according to Appendix A.
\STATE Initialization: $\Gamma_k^i = 0$ and $\eta_{k_v}^i = 0, \forall k, v, i \in \left\{1,2 \right\}$.
\FOR {$k = 1$ to $K$}
\FOR {$i = 1$ to $2$}
\STATE  Find $v = \argmax_{\ddot{v} \in \mathcal{V}} \Gamma_{k_{\ddot{v}}}^i$;
\STATE Update $\Gamma_k^i = \Gamma_{k_v}^i$, $\eta_{k_v}^i = 1$ and $\mathcal{V} = \mathcal{V} - \left\{v\right\}$;
\ENDFOR
\ENDFOR
\REPEAT 
\STATE Find $\left\{k, i\right\} = \argmin_{\ddot{k}, \ddot{i} \in \left\{1,2 \right\}} \Gamma_{\ddot{k}}^{\ddot{i}}$;
\STATE Find  $v = \argmax_{\ddot{v} \in \mathcal{V}} \Gamma_{k_{\ddot{v}}}^i$;
\STATE Update $\Gamma_k^i = \Gamma_k^i + \Gamma_{k_v}^i$, $\eta_{k_v}^i = 1$ and $\mathcal{V} = \mathcal{V} - \left\{v\right\}$;
\UNTIL{$\left(\mathcal{V}  \neq \emptyset\right)$}
\STATE \textbf{Output}: ${\bm \eta}_i, i \in \left\{1,2\right\}$.
\end{algorithmic}
\caption{Sub-Optimal Sub-Band Allocation}
\label{GP_AG}
\vspace{-0.1mm}
\end{algorithm}

\vspace{-8mm}

\subsection{Second-Stage: PS Design at the RIS and Power Allocation}

With the obtained ${\bm \eta}_i, i \in \left\{1,2 \right\}$, the goal of maximizing the minimum bidirectional weighted sum-rate across all the sub-bands for the $K$ node pairs is further achieved by refining ${\bm \Psi}$ and ${\bf p}_i, i \in \left\{1,2\right\}$ through an alternating optimization framework as detailed below.

\subsubsection{PS Design}

%

With the obtained ${\bm \eta}_i$ and fixed ${\bf p}_i, i \in \left\{1,2\right\}$, ${\bm \Psi}$ is obtained by solving $\mathcal{P}_1$, which is non-convex with respect to (w.r.t.) ${\bm \Psi}$ due to ${\rm C}_7$. Accordingly, we propose the following methods to obtain ${\bm \Psi}$:

\paragraph{Exact Solution}

We firstly apply SDR to exactly solve for ${\bm \Psi}$. Accordingly, by introducing an auxiliary variable $\varsigma$, and defining ${\bm \Theta} \triangleq \check{\bm \Psi} \check{\bm \Psi}^H$ and $\check{\bm \Psi} \triangleq \begin{bmatrix}
    {\bm \Psi}  \\
    \varsigma 
\end{bmatrix}$, such that ${\bm \Theta} \succeq	 {\bf 0}$ and ${\rm rank}\left({\bm \Theta} \right) = 1$, $\mathcal{P}_1$ w.r.t. ${\bm \Psi}$ is transformed into the following convex semidefinite program (SDP) by ignoring the rank-one constraint:

\vspace{-2mm}

\begin{equation}
\begin{aligned}
& \mathcal{P}_2: && \underset{\left\{{\bm \Theta}\right\}}{\max}\underset{\left\{i \in \{1,2\}\right\}} \min \; \sum_{k=1}^K  \sum_{v = 1}^{V} \frac{\varkappa_k \;  {\eta_{k_v}^i}}{V} \log_2 \left(1 + \bar{\gamma}_{k_v}^i \left({\bm \Theta} \right)\right) \\
& \text{\it s.t.}
& & {\rm C}_8: \left[{\bm \Theta}\right]_{\left(m,m\right)} = 1, \forall m, {\rm C}_9: {\bm \Theta} \succeq {\bf 0}. 
\end{aligned}
\end{equation}
where $\bar{\gamma}_{k_v}^i \left({\bm \Theta} \right) \triangleq \frac{p_{k_v}^i \left({\rm Tr} \left\{{\bf H}_{k_v}^i {\bm \Theta} \right\} + \left|g_{k_v}^i \right|^2\right)}{\sigma^2_{k_v}}$ and ${\bf H}_{k_v}^i \triangleq \begin{bmatrix}
    {\bf h}_{k_v}^i \left({\bf h}_{k_v}^i\right)^H  & {\bf h}_{k_v}^i g_{k_v}^i \\
    \left( g_{k_v}^i\right)^* \left({\bf h}_{k_v}\right)^H & 0
\end{bmatrix}$. $\mathcal{P}_2$ can be optimally solved by existing convex optimization solvers \cite{Zhang2019APB}, which may not lead to a solution satisfying ${\rm rank}\left({\bm \Theta} \right) = 1$. Accordingly,  for ${\rm rank}\left({\bm \Theta} \right) \neq 1$, Gaussian randomization coupled with the projection operation given  in (16) in Appendix A can be leveraged to obtain ${\bm \Psi}$ as in \cite{Zhang2019APB}. The details are omitted for brevity.



\paragraph{Low-Complexity Solution}

To achieve a lower complexity than the above SDR based solution, let $u_{k_v}^{\bar{i}}, \bar{i} \in \left\{1,2\right\}\backslash\left\{i\right\}$ be the receive filter such that the estimated signal on the $v$-th sub-band at ${\rm Node}_k^{\bar{i}}$ in the $i$-th direction, when $\eta_{k_v}^i = 1$, is given by $\hat{s}^i_{k_v} = u_{k_v}^{\bar{i}} y_{k_v}^i, \forall k, v$. Accordingly, the corresponding mean-squared-error (MSE) is given by


\vspace{-2mm}

\begin{equation}
    \begin{aligned}
        \epsilon^i_{k_v} &= \mathbb{E}\left[\left|s^i_{k_v} - \hat{s}^i_{k_v} \right|^2 \right] = 1 + 2\sqrt{p^i_{k_v}}\Re\left\{{\bm \pi}_{k_v}^i {\bm \Psi} - u_{k_v}^{\bar{i}} g_{k_v}^i \right\}\\
        &\hspace{4em}+ p^i_{k_v} \left( {\bm \Psi}^H {\bm \Pi}_{k_v}^i {\bm \Psi}  + \left|u_{k_v}^{\bar{i}} g_{k_v}^i\right|^2 \right) + \sigma_{k_v}^2 \left|u^{\bar{i}}_{k_v} \right|^2,
    \end{aligned}
\end{equation}
where ${\bm \pi}_{k_v}^i \triangleq \sqrt{p_{k_v}^i} \left(g_{k_v}^i\right)^* \left|u_{k_v}^{\bar{i}} \right|^2 \left({\bf h}^i_{k_v}\right)^H - u_{k_v}^{\bar{i}} \left({\bf h}_{k_v}^i\right)^H$ and
${\bm \Pi}^i_{k_v} \triangleq \left|u_{k_v}^{\bar{i}} \right|^2 {\bf h}_{k_v}^i \left({\bf h}_{k_v}^i\right)^H$. Subsequently, defining ${\bf u}_{\bar{i}} \triangleq \left[u_{1_1}^{\bar{i}}, \cdots, u_{K_V}^{\bar{i}}\right]^T$, we formulate the following problem

\vspace{-2mm}

\begin{equation}
\begin{aligned}
& \mathcal{P}_3: && \underset{\left\{{\bm \Psi}, {\bf w}_1, {\bf u}_1, {\bf w}_2, {\bf u}_2 \right\}}{\min}\underset{\left\{i \in \{1,2\}\right\}} \max \; \sum_{k=1}^K \sum_{v = 1}^{V} \varkappa_k \; \zeta^i_{k_v}\; \text{\it s.t.} \; {\rm C}_7, 
\end{aligned}
\end{equation}
where $\zeta^i_{k_v} \triangleq \frac{\eta_{k_v}^i}{V} \left(w^i_{k_v} \epsilon^i_{k_v} - \log_2 \left(w^i_{k_v} \right) -  1\right)$, $w_{k_v}^i$ is the weight associated with $\epsilon_{k_v}^i$ and  ${\bf w}_i \triangleq \left[w_{1_1}^i, \cdots,  w_{K_V}^i \right]^T$. Given $\mathcal{P}_3$ is difficult to solve due to the coupling of  the variables in $\zeta^i_{k_v}, \forall k, v, i \in \left\{1,2\right\}$, we adopt an inner second-phase alternating optimization to solve $\mathcal{P}_3$. Accordingly, at the $t$-th iteration of the inner alternating optimization, each element of the optimal receive filter ${\bf u}_{\bar{i}}^{(t)}, {\bar{i}} \in \left\{1,2 \right\}\backslash \left\{i\right\}$, for given values of $\left\{{\bf w}_i^{(t-1)}, {\bm \Psi}^{(t-1)}\right\}, i \in \left\{1,2 \right\}$, is equivalent to minimizing $\epsilon^i_{k_v}$ w.r.t. $\left(u^{\bar{i}}_{k_v}\right)^{(t)}$, which is given by

%

\vspace{-2mm}

\begin{equation}
\begin{split}
    \left(u^{\bar{i}}_{k_v}\right)^{(t)} = \sqrt{p^i_{k_v}} \left[p^i_{k_v} \left|{\Xi^{(t-1)}}\right|^2 + \sigma_{k_v}^2 \right]^{-1}  {\Xi^{(t-1)}},    
\end{split}
\end{equation}
where $\Xi^{(t-1)} \triangleq \left(\left({\bm \Psi}^{(t-1)}\right)^H {\bf h}_{k_v}^i + \left(g_{k_v}^i\right)^*\right)$. Subsequently, each element of ${\bf w}_i^{(t)}, i\in \left\{1,2 \right\}$, for the obtained values of ${\bf u}_{\bar{i}}^{(t)}, {\bar{i}}\in \left\{1,2 \right\}\backslash \left\{i\right\}$ and given  ${\bm \Psi}^{(t-1)}$ is computed by minimizing $\mathcal{P}_3$ w.r.t. $\left(w^i_{k_v}\right)^{(t)}$, given by \cite{luo2011} 

\vspace{-2mm}

\begin{equation}
   \left(w^i_{k_v}\right)^{(t)} = \left(\epsilon^i_{k_v}\right)^{-1}.
\end{equation}

\textbf{Lemma 1}: For a given ${\bm \Psi}^{(t-1)}$, the objective of $\mathcal{P}_1$ and $\mathcal{P}_3$ are equivalent for the optimal values of $\left\{ {\bf u}_{\bar{i}}^{(t)}, {\bf w}_i^{(t)} \right\}, i\in \left\{1,2 \right\}$, given by (5) and (6).

\textit{Proof}: The lemma can be proved by substituting (5) and (6) into the objective of $\mathcal{P}_3$. $\blacksquare$

%

Finally, we resort to the PSG method to obtain a low-complexity update for ${\bm \Psi}$. Accordingly, defining $f_i\left({\bm \Psi}^{(t-1)} \right) \triangleq \sum_{k=1}^K \sum_{v = 1}^{V} \varkappa_k \; \zeta^i_{k_v}\big|_{{\bm \Psi} = {\bm \Psi}^{(t-1)}}, i \in \left\{1,2 \right\}$ and $f\left({\bm \Psi}^{(t-1)} \right) \triangleq \max_{i \in \left\{1,2 \right\}} f_i \left({\bm \Psi}^{(t-1)} \right)$, the sub-differential of the unconstrained $\mathcal{P}_3$ for the $t$-th iteration is expressed as $\partial f\left({\bm \Psi}^{(t)} \right) = \operatorname{conv}\left(\bigcup_{i: f_{i}({\bm \Psi}^{(t-1)})=f({\bm \Psi}^{(t-1)})} \nabla f_{i}({\bm \Psi}^{(t-1)})\right)$ \cite{boydSubgradientMethod}, which is the convex hull of the union of gradients of  $f_i \left({\bm \Psi}^{(t-1)} \right), i \in \left\{1,2 \right\}$, that achieve the maximum at ${\bm \Psi}^{(t-1)}$, where


\vspace{-6mm}

\begin{equation*}
\begin{split}
    &\nabla f_i \left({\bm \Psi}^{(t-1)} \right) \triangleq \sum_{k=1}^K \sum_{v=1}^V \eta^i_{k_v} w^i_{k_v} \left({\bf h}^i_{k_v} \left(p^i_{k_v} \left|\left(u^{\bar{i}}_{k_v}\right)^{(t)} \right|^2 \right. \right.\\ 
    &\left.\left.\hspace{0.25em} \times \left(g^i_{k_v} + \left({\bf h}^i_{k_v}\right)^H {\bm \Psi}^{(t-1)} \right) - \sqrt{p^i_{k_v}} \left(\left(u^{\bar{i}}_{k_v}\right)^{(t)} \right)^* \right)\right). \vspace{-8mm}
\end{split}  
\end{equation*}
%
%
Let ${\bm \delta}_{\bm \Psi}^{(t)} \in \partial f\left({\bm \Psi}^{(t)} \right)$ denote any sub-gradient of $f\left({\bm \Psi}^{(t-1)} \right)$ at the $t$-th iteration, where ${\bm \delta}^{(t)}_{\bm \Psi}$ is uniquely given by ${\bm \delta}^{(t)}_{\bm \Psi} = \nabla f_{\Breve{i}} \left( {\bm \Psi}^{(t-1)}\right)$ such that $\Breve{i} = \argmax_{i} f_i\left({\bm \Psi}^{(t-1)} \right)$, when $f_1 \left({\bm \Psi}^{(t-1)} \right) \neq f_2 \left({\bm \Psi}^{(t-1)} \right)$ and ${\bm \delta}_{\bm \Psi}^{(t)} = \tau \nabla f_1 \left( {\bm \Psi}^{(t-1)}\right) +  \left(1 - \tau \right)\nabla f_2 \left( {\bm \Psi}^{(t-1)}\right), \tau \in \left[0,1\right]$, otherwise \cite{boydSubgradientMethod}. Accordingly, ${\bm \Psi}$ is updated as following \cite{andreasson2005introduction}:

\vspace{-2mm}

\begin{equation}
    {\bm \Psi}^{(t)} = {\rm Proj}_{\mathcal{R}} \left({\bm \Psi}^{(t-1)} - \kappa_t {\bm \delta}^{(t)}_{\bm \Psi}/\norm{{\bm \delta}^{(t)}_{\bm \Psi} }_2 \right), \vspace{-2mm}
\end{equation} 
where ${\rm Proj}_{\mathcal{R}} \left(\cdot\right)$ is defined as in (16) in Appendix A, $\kappa_t \triangleq 1/t > 0$ is the diminishing  step size \cite{andreasson2005introduction, boydSubgradientMethod}. Since the PSG method is generally not a decent method, the best value for ${\bm \Psi}$ is given by ${\bm \Psi}^{\rm best} = \argmin_{t = 1, \cdots, T_{max}} f \left({\bm \Psi}^{(t)} \right)$, where $T_{max} \approx 100$ is sufficient to obtain an adequate performance \cite{boydSubgradientMethod}. Algorithm 2 summarizes the proposed framework to obtain  ${\bm \Psi}$ through the PSG method, where $\breve{\bm \Psi}$ denotes the PS vector obtained at the previous iteration of the outer alternating optimization framework. Note that, considering only the dominant computations, the overall complexity of Algorithm 2 is $\mathcal{O}\left(2 T_{max} K V R^2 \right)$, which is significantly less compared to $\mathcal{O}\left(\left(R + 1\right)^6 \right)$ incurred by solving $\mathcal{P}_2$, especially for a large $R$. 



\vspace{-2mm}

\begin{algorithm}[!htb]
 \begin{algorithmic}[1]
\STATE \textbf{Input}:  ${\bf p}_i, i \in \left\{1,2 \right\}$, $\breve{\bm \Psi}$; 
\STATE Initialize ${\bm \Psi}^{(0)} = \breve{\bm \Psi}$;
\FOR {$t = 1$ to $T_{\rm max}$}
\STATE Update ${\bf u}_{\bar{i}}^{(t)}, {\bar{i}}\in \left\{1,2 \right\}\backslash \left\{i\right\}$ using (5);
\STATE Update ${\bf w}_i^{(t)}, i \in \left\{1,2\right\}$ using (6);
\STATE Update ${\bf \Psi}^{(t)}$ using (7);
\ENDFOR
\STATE \textbf{Output}: ${\bm \Psi} = {\bm \Psi}^{\rm best}$.
\end{algorithmic}
\caption{PS Design with PSG Method}
\label{GP_AG}
\vspace{-0.35em}
\end{algorithm}

\vspace{-2.25mm}

\subsubsection{Power Allocation}

%


For the obtained ${\bm \eta}_i, i \in \left\{1,2\right\}$ and ${\bm \Psi}$, ${\bf p}_i, i \in \left\{1,2\right\}$ is computed by solving $\mathcal{P}_1$, which has a waterfilling solution, given by $p_{k_v}^i  = \left[\frac{1}{\sum_{v=1}^V \eta_{k_v}^i} \left(P^i_k + \sum_{v=1}^V \frac{\eta_{k_v}^i}{\varpi_{k_v}^i} \right) - \frac{1}{\varpi_{k_v}^i} \right]^+$ if $\eta^i_{k_v} = 1$ and $p_{k_v}^i  = 0$, otherwise, where $[q]^+ \triangleq \max(0,q)$, $\varpi_{k_v}^i \triangleq \frac{\left|g_{k_v}^i + \left({\bf h}_{k_v}^i\right)^H {\bm \Psi} \right|^2}{ \sigma_{k_v}^2}$. Furthermore, to efficiently utilize the total transmit power of each node, we adopt the iterative waterfilling algorithm, similar to that described in \cite{cioffi2006}. Specifically, in each iteration, we set $\eta_{k_{\bar{v}}}^i = 0$ such that $\bar{v} = \argmin_{\Ddot{v}} \varpi_{k_{\Ddot{v}}}^i$, if the obtained $p_{k_{\bar{v}}}^i = 0$ when $\eta_{k_{\bar{v}}}^i = 1$, where the algorithm continues till $p_{k_v}^i > 0, \forall \; \eta_{k_v}^i = 1$. Note that the details are avoided for brevity. Finally, Algorithm 3 summarizes the overall framework to maximize the minimum bidirectional weighted sum-rate for the proposed two-way communication.


\setlength{\textfloatsep}{10pt}

\begin{algorithm}[!htb]
 \begin{algorithmic}[1]
\STATE \textbf{Input}: $g_{k_v}^i$, ${\bf h}_{k_v}^i, \forall k, v, i \in \left\{1,2 \right\}$.
\STATE Initialize ${\bm \Psi}$ according to Appendix A;
\STATE Obtain ${\bm \eta}_i, i \in \left\{1,2 \right\}$ using Algorithm 1;
\REPEAT
\STATE Update ${\bm \Psi}$ by solving $\mathcal{P}_2$ or using Algorithm 2;
\STATE Update ${\bf p}_i, i \in \left\{1,2\right\}$ using iterative waterfilling;
\UNTIL convergence
\STATE \textbf{Output}: ${\bm \eta}_i$, ${\bf p}_i, i \in \left\{1,2 \right\}$, ${\bm \Psi}$.
\end{algorithmic}
\caption{Proposed Two-Way Communication Algorithm}
\label{GP_AG}
\vspace{-0.35em}
\end{algorithm}

\vspace{-2.19mm}

\section{Numerical Results}  

\vspace{-0.75mm}

In  this  section,  we  evaluate  the  performance of our proposed design via Monte-Carlo simulations. Unless stated otherwise, we assume $K = 3$,  $V = 16$, $P_k^i = 25 \; {\rm dBm}$ and $\sigma_{k_v}^2 = -110 \; {\rm dBm}, \forall k, v, i \in \left\{1,2\right\}$ \cite{ruiZhangOfdma2020}. Considering the Cartesian coordinate system, the RIS is located at $\left(0,0,10\;m \right)$, where ${\rm Node}_k^1$ and ${\rm Node}_k^2, \forall k$ are uniformly distributed within a sphere of radius $5\; {\rm m}$ centered around $\left(-35\; {\rm m},0,5 \; {\rm m}\right)$ and $\left(35\; {\rm m},0,5 \; {\rm m} \right)$, respectively. The sum-rate weights are set as $\varkappa_k = 1, \forall k$. The path loss of each channel is modeled by $\varrho_x = \varrho_0 \left(d_x/d_0 \right)^{-\beta_x}, x \in \left\{k-k, k-r, r-k \right\}$, where $\varrho_0 = -30 \; {\rm dB}$ denotes the reference path loss at the reference distance $d_0 = 1 \; {\rm m}$, $d_x$ and $\beta_x$ are the link distance and the path loss exponent, respectively. The path loss exponents are set as $\beta_{k-k}= 3.5$ and $\beta_{k-r} = \beta_{r-k} = 2.2, \forall k, r$.  For each multi-path channel,  the maximum delay tap is set as $L_{k-k} = 8$ and $L_{k-r} = L_{r-k} = 4, \forall k, r$, and $\alpha = 0.5$ \cite{ruiZhangOfdma2020, analysis2008}. The parameters for Algorithn 2 is set as $\tau = 0.5$ and $T_{max} = 100$. For clarity, the following schemes, which are various instances of Algorithm \ref{GP_AG}, and the corresponding abbreviations are used throughout this section: 1) \textit{optSDR/optPSG:} Proposed designs based on the SDR and PSG methods; 2) \textit{uniPowPSG:} Proposed design based on the low-complexity PSG method with a uniform power allocation across the allocated sub-bands; 3) \textit{initialPSs:} Proposed design where ${\bm \Psi}$ is fixed to $\bar{\bm \Psi}$ as designed in Appendix A; 4) \textit{randInitialPSG:} Proposed design based on the PSG method where ${\bm \Psi}$ is randomly initialized; 5) \textit{randPSs:} Proposed design where ${\bm \Psi}$ is randomly generated, and 6) \textit{noRIS:} Proposed design with only direct link between ${\rm Node}_k^1$ and ${\rm Node}_k^2, \forall k$.



\vspace{-4.5mm}

\begin{figure}[!htb]
 \centering
   \subfigure[ ]
    {
        \includegraphics[scale=0.185]{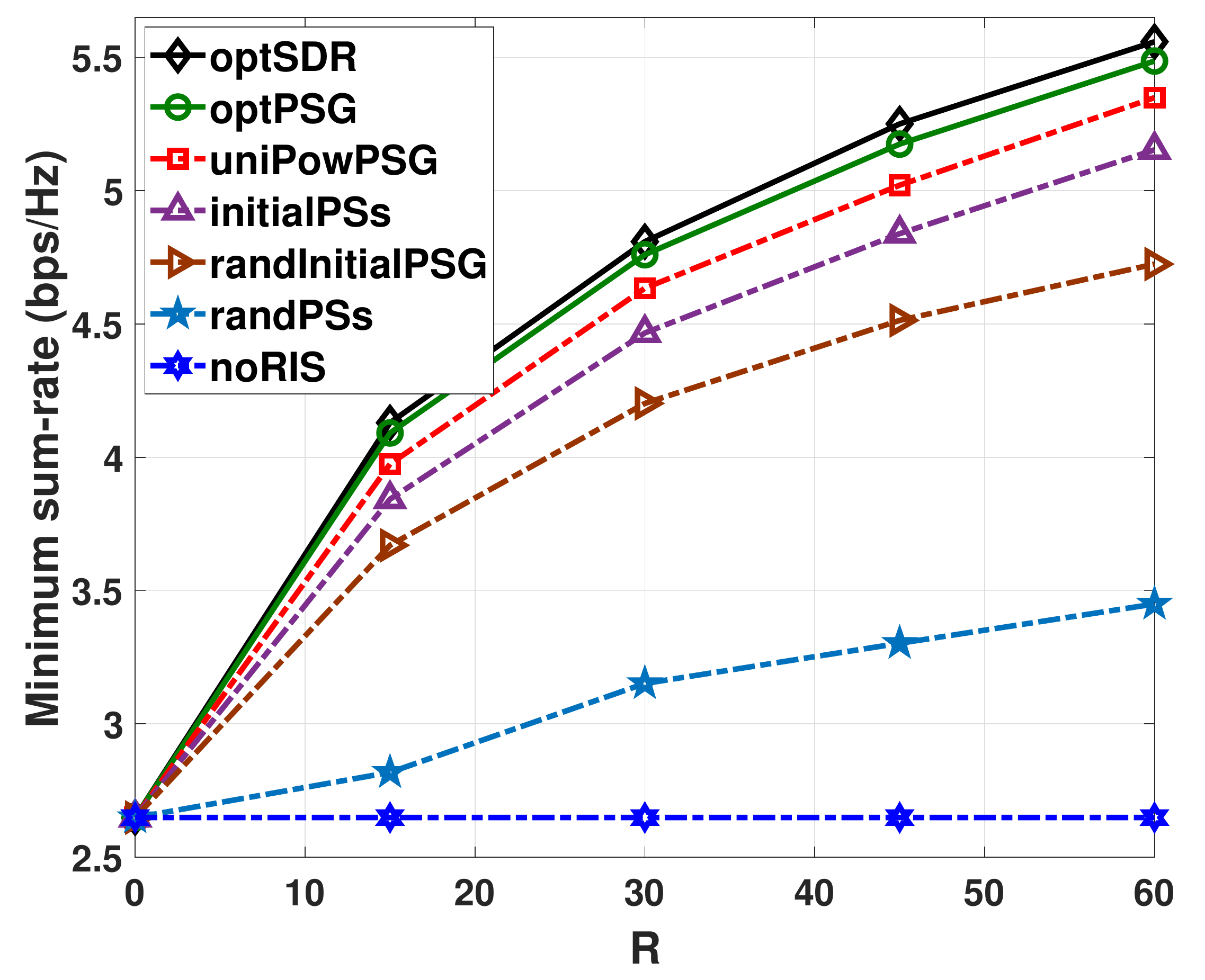}
    }  \hskip -2.75ex
    \subfigure[] 
    {
         \includegraphics[scale=0.185]{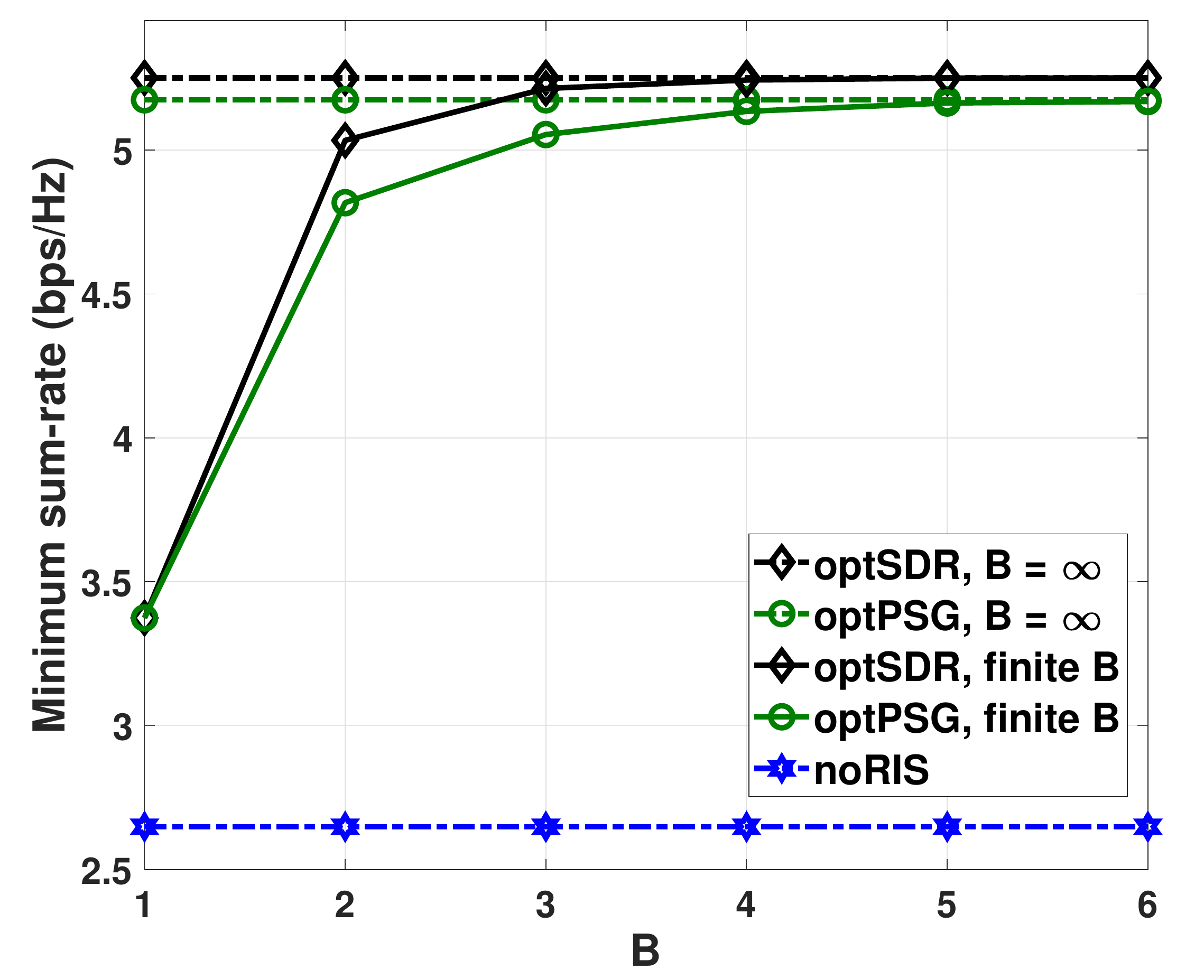}
 }     \vspace{-3.25mm}
    \caption{\small Minimum sum-rate v.s. a) $R$ with $B = \infty$, b) $B$ with $R = 45$ for $K = 3$, $V = 16$, $P_k^i = 25\; {\rm dBm}, \forall k, i \in \left\{1,2 \right\}$.}
    \vspace{-3.0mm}
\end{figure}

 Firstly, the minimum sum-rate performance among both the directions w.r.t. $R$, assuming $B = \infty$, is shown in Fig. 2(a), where the performance gain is observed to improve with an increase in $R$. Note that the low-complexity optPSG design incurs a marginal performance loss compared to the optSDR design. Furthermore, the higher performance of optPSG design compared to the other designs establishes the merit of the proposed initialization for ${\bm \Psi}$ and the proposed update for $\left\{{\bf p}_i, {\bm \Psi}\right\}, i \in \left\{1,2\right\}$ as described in Algorithm 3.  Finally, in Fig. 2(b), the minimum sum-rate performance for the proposed optSDR and optPSG designs is evaluated w.r.t. the finite value of $B$, where the proposed designs with $B = 5$ are seen to attain an indistinguishable performance to the case when $B = \infty$. Moreover, it can be observed that even with $B = 1$, the proposed designs achieve a higher performance compared to the case where there is no deployment of the RIS. 
  
 

\vspace{-2mm}
\section{Conclusion}

In this paper, we have maximized the minimum bidirectional weighted sum-rate for a RIS-enhanced two-way D2D OFDM communication system with multiple node pairs by jointly optimizing the sub-band allocation, the power allocation and the discrete PS design at the RIS. The main challenge was to design the PSs at the RIS, which was obtained through a SDR formulation and an equivalent low-complexity solution based on the PSG method. The desired performance gain of the proposed designs has been validated through numerical examples. 

\vspace{-4.0mm}

\appendix

\vspace{-1.0mm}

  \subsection{Initialization for ${\bm \Psi}$}
  
    \label{FirstAppendix}

  The constant-modulus constraint on ${\bm \Psi}$, i.e., ${\rm C}_7$, makes $\mathcal{P}_1$ highly non-convex, resulting in multiple local minimum points for $\mathcal{P}_1$. Accordingly, to ensure that the proposed algorithm converges to a near-optimum local point, initialization for ${\bm \Psi }$ plays a crucial role. For this purpose, inspired by the work in \cite{Zhang2019APB}, and defining ${\bf g}_i \triangleq \left[g_{1_1}^i, \cdots, g_{K_V}^i  \right]^T \in \mathbb{C}^{KV \times 1}$ and ${\bf H}_i \triangleq \left[{\bf h}^i_{1_1}, \cdots, {\bf h}_{K_V}^i \right]^H \in \mathbb{C}^{KV \times R}$, we initialize ${\bm \Psi}$ to the solution of the following optimization problem:

%
  
  \vspace{-1mm}

  \begin{equation}
\begin{aligned}
& \mathcal{P}_4: && \underset{\left\{{\bm \Psi} \right\}}{\max}\underset{\left\{i \in \{1,2\}\right\}} \min \; \norm{{\bf g}_i + {\bf H}_i {\bf \Psi}}_2^2 \; \text{\it s.t.} \; {\rm C}_7,
\end{aligned}
 \vspace{-2mm}
\end{equation}
which maximizes the minimum bidirectional effective channel gain across all the sub-bands of the $K$ node pairs. Accordingly, relaxing ${\rm C}_7$ and ignoring the terms independent of ${\bm \Psi}$, $\mathcal{P}_4$ can be transformed into the following epigraph form:

\vspace{-1mm}

\begin{equation}
\begin{aligned}
\mathcal{P}_5: \underset{\left\{\tilde{\bm \Psi}, \rho \right\}}{\max} \; \rho \; \; \text{\it s.t.} \; &{\rm C}_{10}: {\rm Tr}\left\{\tilde{\bm \Psi}^H \tilde{\bf H}_i \tilde{\bm \Psi} \right\} \geq \rho, i \in \left\{1,2 \right\}, \\
&{\rm C}_{11}: \norm{\tilde{\bf \Psi}}_2^2 \leq R+1,
\end{aligned}
\vspace{-2mm}
\end{equation}
%
%
where $\tilde{\bm \Psi} \triangleq \begin{bmatrix}
    {\bm \Psi}  \\
    1 
\end{bmatrix}$ and $\tilde{\bf H}_i \triangleq \begin{bmatrix}
    {\bf H}_i^H {\bf H}_i  & {\bf H}_i^H {\bf g}_i \\
    {\bf g}_i^H {\bf H}_i & 0
\end{bmatrix}$. Subsequently, the Lagrangian associated with $\mathcal{P}_5$ is given by

\vspace{-2mm}

\begin{equation}
    \mathcal{L} = -\rho + \sum_{i = 1}^2\lambda_i \left(\rho -   {\rm Tr}\left\{\tilde{\bm \Psi}^H \tilde{\bf H}_i \tilde{\bm \Psi} \right\} \right) + \mu \left(\norm{\tilde{\bf \Psi}}_2^2 - R -1\right), \vspace{-2mm}
\end{equation}
and the corresponding the KKT conditions are given by

\vspace{-2mm}

    \begin{align}
\pdv{\mathcal{L}}{\tilde{\bm \Psi}^H}  =  - \sum_{i = 1}^2 \lambda_i   \tilde{\bf H}_i \tilde{\bm \Psi} + \mu \tilde{\bm \Psi} = 0,&\\
\pdv{\mathcal{L}}{\rho}  = -1 + \sum_{i=1}^2 \lambda_i = 0, & \\
\lambda_i \left(\rho -   {\rm tr}\left\{\tilde{\bm \Psi}^H \tilde{\bf H}_i \tilde{\bm \Psi} \right\}\right)  = 0, \; \lambda_i \geq 0, i \in \left\{1,2\right\}, &\\
\mu \left(\norm{\tilde{\bf \Psi}}_2^2 - R - 1\right) = 0, \; \mu \geq 0,
\end{align}
%
where $\left\{\lambda_i, \mu \right\}, i \in \left\{1,2\right\}$ are the Lagrangian multipliers. Note that, from (12), we have $\sum_{i = 1}^2\lambda_i = 1$, which implies $0 \leq \lambda_i \leq 1, i \in \left\{1, 2\right\}$. Accordingly, leveraging (11), we obtain the following condition:

\vspace{-2mm}

\begin{equation}
\begin{split}
        &\left(\lambda_1 \tilde{\bf H}_1 + \left(1 - \lambda_1\right)\tilde{\bf H}_2\right){\tilde{\bm \Psi}} = \mu \tilde{\bm \Psi},\\
        \implies & \tilde{\bf H} \left(\lambda_1\right) \tilde{\bm \Psi} = \mu \tilde{\bm \Psi},\\
\end{split}
\end{equation}
where $\tilde{\bf H} \left(\lambda_1\right)  \triangleq \left(\tilde{\bf H}_2 + \lambda_1\left(\tilde{\bf H}_1 - \tilde{\bf H}_2\right)\right)$ is a Hermitian matrix. Subsequently, by expressing the eigen decomposition of $\tilde{\bf H} \left(\lambda_1\right)$ as $\tilde{\bf H} \left(\lambda_1\right) \triangleq {\bf U}_{\lambda_1} {\bm \Sigma}_{\lambda_1} {\bf U}_{\lambda_1}^H$, the  solution to $\mathcal{P}_5$ as a function of $\lambda_1$ is given by the principal eigenvector corresponding to the maximum eigenvalue of $\tilde{\bf H} \left(\lambda_1\right)$, denoted by $\tilde{\bf u}_{\lambda_1}$. Subsequently, the optimal initial ${\bm \Psi}$ which maximizes the minimum bidirectional effective channel gain across all the sub-bands of the $K$ node pairs, while satisfying ${\rm C}_7$, is given by

\vspace{-4mm}

\begin{equation}
    \bar{\bm \Psi} =  {\rm Proj}_{\mathcal{R}} \left(\grave{\bm \Psi} \right).
    \vspace{-2mm}
\end{equation}
where ${\rm Proj}_{\mathcal{R}} \left(\grave{\bm \Psi} \right) \triangleq \argmin_{\Ddot{\bm \Psi}\in \mathcal{R}} \abs{\Ddot{\bm \Psi} - \grave{\bm \Psi}}$, $\grave{\bm \Psi} \triangleq \frac{\left[\tilde{\bf u}_{\bar{\lambda}_1}\right]_{(1:R)}}{\left[\tilde{\bf u}_{\bar{\lambda}_1}\right]_{(R+1)}}$, $\bar{\lambda}_1 \triangleq \argmax_{\lambda_1} \min_{\left\{i \in \{1,2\}\right\}} {\rm Tr}\left\{\tilde{\bf u}_{\lambda_1}^H \tilde{\bf H}_i \tilde{\bf u}_{\lambda_1} \right\}$ and the optimal $\bar{\lambda}_1$ can be obtained through a linear search over $0 \leq \lambda_1 \leq 1$. 


\ifCLASSOPTIONcaptionsoff
  \newpage
\fi

\vspace{-5.75mm}

\bibliography{myBib.bib}
\bibliographystyle{IEEEtran}

\end{document}